\documentclass[preprint,12pt]{elsarticle}

\usepackage{graphicx}
\usepackage{tikz}
\usetikzlibrary{positioning,shapes,fit,arrows}

\definecolor{myblue}{RGB}{56,94,141}
\usepackage{amssymb}
\usepackage{amsthm}
\usepackage{amsmath}

\theoremstyle{definition}
\newtheorem{definition}{Definition}

\theoremstyle{definition}

\theoremstyle{remark}
\newtheorem{assumption}{\textbf{Assumption}}

\theoremstyle{remark}
\newtheorem*{remark}{Note}

\usepackage{lineno}

\journal{arXiv}

\begin{document}

\begin{frontmatter}


\title{A Matching Based Theoretical Framework for Estimating Probability of Causation}



\author{Tapajit Dey and Audris Mockus}
\address{Department of Electrical Engineering and Computer Science\\ University of Tennessee, Knoxville}
\ead{tdey2@vols.utk.edu, audris@utk.edu}

\begin{abstract}
  The concept of Probability of Causation (PC) is critically important
  in legal contexts and can help in many other domains. While it has
  been around since 1986, current operationalizations can obtain
  only the minimum and maximum values of PC, and do not apply for
  purely observational data. We present a theoretical framework to
  estimate the distribution of PC from experimental and from purely
  observational data. We illustrate additional problems of the
  existing operationalizations and show how our method can be used to address
  them. We also provide two illustrative examples of how our method
  is used and how factors like sample size or rarity of
  events can influence the distribution of PC. We hope this will
  make the concept of PC more widely usable in practice.
\end{abstract}


\end{frontmatter}


\section{Introduction}\label{s:intro}
Our understanding of the world comes from our knowledge about the cause-effect relationship between various events.  However, in most of the real world scenarios, one event is caused by multiple other events, having varied degrees of influence over the first event. There are two major  concepts associated with such relationships: Cause of Effect (CoE) and Effect of Cause (EoC)~\cite{dawid2016statistical,pearl2015causes}. EoC focuses on the question that if event X is the cause/one of the causes of event Y, then what is the probability of event Y given event X is observed? CoE, on the other hand, tries to answer the question that if event X is known to be (one of) the cause(s) of event Y, then given we have observed both events X and Y, what is the probability that event X was in fact the cause of event Y? 

In this paper, we focus on the CoE scenario. A measure of the
probability for CoE situations was offered by Tian and
Pearl~\cite{tian2000probabilities}, which they call probability of
necessity (PN); and also by Philip
Dawid~\cite{dawid2016statistical}, who call it Probability of
Causation. The definition of Probability of Causation/Necessity (in
our paper we call it Probability of Causation - PC) is based on a
counterfactual definition that is used by both
Dawid~\cite{dawid2016statistical,dawid2016bounding}, and
Pearl~\cite{tian2000probabilities}.  If X and Y are assumed to be
binary valued events, the definition of PC then states that given we
have observed a positive outcome for both event X and Y, the value
of PC is the probability that had the event A been negative, event B
would also be negative. This definition is counterfactual because it
is assuming a scenario that didn't actually happen. So, the
probability values must be calculated by examining the data from
other known observations involving events X and Y, in an
experimental and/or observational scenario. Generally the data from
an experimental setting is preferred because many of the biases are
accounted for in a carefully done randomized experiment~\cite{rubin1978bayesian}.

The application of concepts related to CoE are most useful in
situations where assigning responsibilities is important, such as in
legal contexts, finding the effects/side-effects of a drug, and
determining risk factors for a disease. However, we believe that
determining the CoE is useful in every scientific discovery that
speculates about one specific event causing another. For example, we
might want to know if the increase in popularity of a particular
technology caused a specific developer to start using it or if a
specific change to the source code can be attributed to a specific
author. In both cases we have both the observational distribution of
events describing frequencies reflecting these relationships and
specific instances (of technology use and of author label attached
to a commit), but we need to ascertain causality (whether the use
of technology was inspired by popularity and whether the specific author
label indicates the actual author of the code change). Another
example from forensic anthropology context might involve the
question if a specific set of weather events caused mummification of
the corpse. Thus we wanted to use
the relevant concepts in practical situations beyond the legal
context. We started from the PN
(PC) formula given by Tian and Pearl~\cite{tian2000probabilities},
because of the claim that it can be used to gain information from
experimental and observational data alike, since in many practical
situations, conducting a randomized experiment may not be possible.  


However, we found that the probability measure given by the PN
formula is very unstable with respect to small changes in the data
(see example in Section~\ref{s:method}). This makes it difficult to
apply the result in practice, because it would be undesirable to
have for example, measurement errors that are common in most
practical cases, have undue influence on the outcomes. In this
paper, therefore, we propose a way to make the probability of
causation be more robust with respect to small changes in the
data. We were also looking for a more simple and intuitive way (than
solving a set of linear equations) of obtaining the formula for the
probability of causation, to make its meaning more understandable to
help explain it to people outside the scientific community. In
Section~\ref{s:theory}, we will present an alternate way of
arriving at the PC formula, given by Tian and
Pearl~\cite{tian2000probabilities}, and
Dawid~\cite{dawid2016statistical}, which we believe to be simpler
and more intuitive. The limits for the probability value obtained by
our approach are the same as given in the two above-mentioned
papers, but we also obtain a distribution of values in that interval.

The rest of the paper is organized as follows: in Section~\ref{s:history}, we discuss the advent and application of the
concept of Probability of Causation in previous studies. In
Section~\ref{s:theory}, we present our approach to the
concept. In Section~\ref{s:method}, we highlight the
robustness issues mentioned earlier, and present our proposed
adjustment methods. In Section~\ref{s:lim}, a few limitations of the
concept of Probability of Causation are listed, along with a
limitation particular to the PN
formula~\cite{tian2000probabilities}. Finally, in Section~\ref{s:conclusion}, we discuss the implications of our work and
conclude the paper.  

\section{Brief History}\label{s:history}

The idea of Probability of Causation dates back to Hume (1748) and
Mill (1843) and has been formalized and advocated in the
philosophical work of D. Lewis~\cite{lewis1986probabilities}. Robins
and Greenland~\cite{robins1989probability} gave a mathematical model
for Probability of Causation in their 1989 paper. In his 1999
paper~\cite{pearl1999probabilities}, Pearl has given a detailed
account of different types of probabilities related to the concept
of Probability of Causation, viz. Probability of Necessity (PN -
which by definition is same as PC), Probability of Sufficiency (PS),
and Probability of necessary-and-sufficient causation (PNS). In the
paper published in 2000~\cite{tian2000probabilities}, Tian and Pearl
have expanded on the previous work and has given the upper and lower
bounds under various conditions. While Pearl approached the problem
from a structural equation modeling perspective,
Dawid~\cite{dawid2000causal} approached the problems from concept
similar to a joint distribution over the different cause-effect
situations. In his later
works~\cite{dawid2016statistical,dawid2016bounding}, Dawid has
worked on defining the limits of PC under various confounding and
mediation scenarios. Here we are not concerned with the
philosophical aspects of causation in objective reality as these, as
clearly demonstrated by Kant~\cite{kant1999critique} can never
be ascertained with certainty, but with
issues arising when applying the formalisms developed in this domain
to practical problems. 

\section{Proposed Theoretical Interpretation of Probability of Causation}\label{s:theory}

\section*{Limitations of current approaches to Probability of Causation}

The current approaches to calculating Probability of Causation, as
illustrated by Pearl and Dawid, have a few limitations, both from
the conceptual and application perspectives, as listed below:
\begin{itemize}
\item \textbf{The counterfactual definition:} The first question
  related to the notion of PC is how the probability is defined in
  this case. In general, probability is defined as the likelihood of
  occurrence of an event; but since the counterfactual definition
  talks about a fictitious event in a fictitious reality that never 
  occurred, the general definition can not be
  applied in this context, giving rise to doubts if it is indeed a
  probability or just some function bounded between 0 and 1. Due to
  the inapplicability of the likelihood of event based definition in
  this context, many of the common concepts associated with the
  general definition of probability are hard to define here, such as
  probability density function, expectation etc. 
Moreover, the mechanism by which the transition between the two
realities occur is not clear in the current approaches. 

We also use the counterfactual definition in our method, but our way
of defining the mechanism makes sure that the measurement of PC is
based on real events. 

\item \textbf{Need for Experimental data:} The current approaches 
  need to have data from a controlled experiment to be able to estimate
  the value of PC. The methodology offered by Tian and 
  Pearl~\cite{tian2000probabilities} can be used to estimate PC from 
  observational data when experimental data for the same situation is
  also available. However, in practical situations, conducting controlled
  experiments is often very difficult, if not impossible, limiting the 
  applicability of the concept.

\item \textbf{No Distribution of PC:} The current approaches focus
  on identifying the upper and lower bounds of PC, but offer no
  description of how the distribution of PC would look like, which
  makes it difficult to apply concepts of expectation and calculate
  variance over the quantity, which would be very important for a
  practical application of the concepts. In Dawid's approach,
  although it is possible to get a posterior distribution for PC by
  putting a custom prior, the authors themselves admit the
  assumptions to be ``highly unrealistic", and therefore say that ``
  ... our analysis must not be taken as delivering a credible
  conclusion ..."~\cite{dawid2016statistical}.  

\item \textbf{Source of Probabilistic Uncertainty not defined:} A
  concept related to the point above, probabilistic uncertainty for
  the random variables in question is essential in defining the
  distribution for PC. Pearl arrived at the upper and lower bounds
  of PN by using structural equation modeling, and it is not evident
  from the formulation where the uncertainties come into picture. In
  Dawid's case also, it is not directly clear where the
  uncertainties are introduced in the equation other than by
  assuming a custom prior distribution for a variable.  

\item \textbf{The G2i problem:} The problem of how to apply the
  result found by experimenting on a certain group can be applied to
  an individual case (the Group to Individual, or G2i problem) was
  mentioned by Dawid in \cite{dawid2015causes}. However, no
  particular way to address the problem was discussed, except from
  an assumption of exchangeability, which states the individual(s)
  under consideration can be exchanged with any participant in the
  experiment, without altering the findings. However, in most of the
  real life situations, this assumption may not be valid.  
  
\item \textbf{Sensitivity to Data Errors:} Being precise mathematical 
  definitions, the current method of estimating PC is extremely sensitive
  to small data errors for rare events, as highlighted in 
  Section~\ref{s:method}. However, small errors in data are very common 
  in practical scenarios. Thus, the result of the current approaches in 
  such situations can be completely inaccurate. 


\end{itemize}

\section*{Proposed Theoretical Framework }

In this section we present our approach for deriving the Probability
of Causation formula. Since we are dealing with two alternate realities 
as a part of defining the concept of Probability of Causation, we need the
observations across the two realities to have the same distribution, and 
also the same property in the sense that they are affected by the cause event 
identically. Moreover, the observations should be exchangeable within each 
reality. This assumption is a precondition of any experimental setting, but if
these assumptions are satisfied, our theoretical framework can be used to 
estimate the value of PC from purely observational scenarios as well.

Let us list the assumptions we have in place for the operationalization of the 
definition of PC.  
\begin{assumption}
X and Y are two binary valued events.
\end{assumption}
\begin{assumption}
X is known to be a cause of Y.
\end{assumption}
\begin{assumption}
$\mathbb{A}$, $\mathbb{B}$, $\mathbb{C}$, and $\mathbb{D}$ are four mutually exclusive sets. The definition for these sets is given later, and is also illustrated in Figure \ref{fig:mapping}.
\end{assumption}
\begin{assumption}
$|\mathbb{A}| + |\mathbb{B}| = |\mathbb{C}| + |\mathbb{D}| = N; $ ($|.|$ represents the cardinality of a set).
\end{assumption}



\begin{assumption}
X = 0 and X = 1 represent two alternate realities (called $X_0$ and $X_1$ hereafter, respectively), as can be seen from Figure \ref{fig:mapping}.
\end{assumption}

\begin{definition}
Under these assumptions, we define Set $\mathbb{A}$ as the set of all elements for which event Y takes a value of 0 under the reality of $X_0$. Sets $\mathbb{B}$, $\mathbb{C}$, and $\mathbb{D}$ are defined similarly. 
\end{definition} 

Under the counterfactual definition of PC, as was used by Pearl and Dawid, for an element $z \in \{\mathbb{A}, \mathbb{B}, \mathbb{C}, \mathbb{D}\}$, we will have 
\begin{align*}
PC = P (z \in \mathbb{A} | z \in \mathbb{D}), && \text{where P(.) indicates probability }
\end{align*}

I.e., given an element $z$ had the value of the event Y to be 1 under the reality of $X_1$,  probability of causation is the probability of same element $z$ having the value of the event Y to be 0 under the reality of $X_0$. Dawid's approach does not explicitly differentiates between the two realities, and derives the formula for PC from a joint distribution over the the two realities. Pearl recognizes the two realities as different, and defines the outcomes under different \textit{do} operators, however, how the \textit{do} operation can be used to move from one reality to another, and under which specific cases such a transition is possible for an already observed case, remains unclear.

\section*{Proposed Mechanism for calculating PC using two realities:}

\begin{figure}
\includegraphics[width = \linewidth]{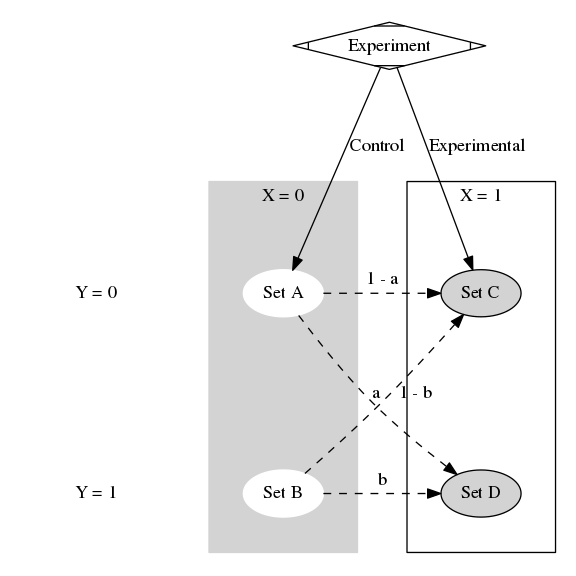}
\caption{Proposed mapping framework between the different sets in the context of a typical experimental scenario}
\label{fig:mapping}
\end{figure}

We define our proposed mechanism for transition between the two realities, viz. $X_0$ and $X_1$, is based on the mapping between the Sets  $\mathbb{A}$, $\mathbb{B}$, $\mathbb{C}$, and $\mathbb{D}$. An element $z$ can be in either set $\mathbb{A}$ or $\mathbb{B}$ in $X_0$, and in set $\mathbb{C}$ or $\mathbb{D}$ in $X_1$. An element can go from set $\mathbb{A}$ or $\mathbb{B}$ to set $\mathbb{C}$ or $\mathbb{D}$ only when the value of X changes, and all elements from sets $\mathbb{A}$ and $\mathbb{B}$ go to either set $\mathbb{C}$ or $\mathbb{D}$ when the value of X changes (Assumption 4) . This mapping is illustrated in Figure \ref{fig:mapping} in the context of a typical experimental situation. Without loss of generality we can assume that the condition X = 0 was enforced upon the control group in an experimental setting, making it a part of the reality $X_0$, and the condition X = 1 was enforced upon the experimental group in the experiment, making it a part of the reality $X_1$. 

Now, in practice, it is almost impossible to emulate the exact experimental conditions and change only the value of the precondition X, which would be the situation required to obtain an exact mapping. So, instead, for every element $z : z \in \{\mathbb{A}, \mathbb{B}\}$, we find an element $z' : z ' \in \{\mathbb{C}, \mathbb{D}\}$, such that $z$ is most similar to $z'$. The information on other attributes of the elements (such as age, gender, race etc. for a patient in a hypothetical clinical trial) can be used to find the best matching pairs. Then we define the mapping between the sets by looking at the mappings between each pair of elements.

For the purpose of defining the probability of causation, all we need from this mapping is what fraction of elements in Set $\mathbb{A}$ go to Set $\mathbb{D}$ and what fraction of elements in Set $\mathbb{B}$ go to Set $\mathbb{D}$, for all elements in Set $\mathbb{D}$. It is also possible to take a set of matching elements instead of just one element for each element in Set $\mathbb{D}$, and thereby instead of having a one-to-one mapping for each element, have a probability of the element having come from one of the two sets, and have the final mapping by averaging the probabilities for each element, with an associated variance. The analysis in this paper is done assuming the one-to-one mapping scenario, however, it should be possible to extend the analysis for a probabilistic mapping scenario without much trouble. 
\begin{remark}
Hereafter, when we say one element moves from one set to another, we essentially mean the second set contains the closest matching element to the element of the first set.
\end{remark}

So, by that relation, we can say:
\begin{equation}\label{eq:1}
|\mathbb{D}| = a \times |\mathbb{A}| + b \times |\mathbb{B}| 
\end{equation}
\begin{equation}\label{eq:2}
|\mathbb{C}| = (1 - a) \times |\mathbb{A}| + (1 - b) \times |\mathbb{B}| 
\end{equation}
where $a$ and $b$ are, respectively,  the fraction of elements from set $\mathbb{A}$ and set $\mathbb{B}$ that go to set $\mathbb{D}$, with $0 \leqslant a,b \leqslant 1$. 
We call the the two fractions $a$ and $b$ \textbf{transition coefficients}, formally defined as:
$$
a = \frac{|\{(z,z'):z' \in \mathbb{D}, z \in \mathbb{A}, f_m(z, z') \geqslant T\}|}{|\{z : z \in \mathbb{A}\}|} 
$$
$$
b = \frac{|\{(z,z'):z' \in \mathbb{D}, z \in \mathbb{B}, f_m(z, z') \geqslant T\}|}{|\{z : z \in \mathbb{B}\}|} 
$$
where, $f_m$ is the matching function used and $T$ is some custom defined threshold.

\section*{Measuring the Value of Probability of Causation}

Using our mechanism, the definition for probability of causation would be: 
\begin{definition}
Probability of Causation: The probability that for an element chosen at random from the set of elements (Set $\mathbb{D}$ in our example) for which the value of the effect event under consideration (event Y in our example) is observed to be 1 (or positive/True) under the reality/precondition  of the cause event ( event X in our example) having a value of 1 (or positive /True), the closest matching element can be found in the set of elements (Set $\mathbb{A}$ in our example) for which the value of the effect event under consideration (event Y in our example) is observed to be 0 (or negative/False) under the reality/precondition  of the cause event ( event X in our example) having a value of 0 (or negative/False).
\end{definition}

So, mathematically, Probability of Causation (PC) would be:
\begin{equation}\label{eq:pc}
PC = \frac{|\{ (z, z') : z \in \mathbb{D}, z' \in \mathbb{A}, f_m(z, z') \geqslant f_m(z, k), \forall k \in \{\mathbb{A}, \mathbb{B} \} \}|}{|\mathbb{D}|}
\end{equation}
where $f_m$ is the matching function used and $(z, z')$ indicate one
pair of matching elements. It should be noted that our formula of PC
does not involve any counterfactual notion, in the sense that we do
not try to calculate what would have been the case for the same
element under a different reality,   instead we try to find the
closest matching element, and calculate the probability from how
that element would behave under the different reality. 

\begin{remark} Addressing the G2i problem:\\
To address the G2i problem, we need to use the exchangeability 
assumption. 
Under the exchangeability assumption we can simply replace any
observed element $z_o$ with an element $z$ from Set $\mathbb{D}$,
and apply the same formula, which takes care of the G2i problem. If
the assumption does not hold for all elements of Set $\mathbb{D}$
however, we can modify the formula slightly to not consider all
elements of Set $\mathbb{D}$, but only a subset of elements from the
set, defined by:  $\{ z'' : f_m(z_o, z'') \geqslant T,  z'' \in
\mathbb{D} \}$; where $T$ is a threshold value of choice. In the
formulas used in this paper, the exchangeability assumption is made,
however, it should not be difficult to modify the formulas using the
constraint if the assumption can not be made for a particular case. 
\end{remark}

\begin{remark}
For convenience of notations, we denote the elements $z$ and $z'$ as the same element $z$. Reader should be mindful that whenever we are talking about transitioning between the realities $X_0$ and $X_1$, we are in fact talking about the closest matching elements.
\end{remark}
 By this convention of notation, we can write the PC formula as:

\begin{equation*}
PC = \frac{|\{z: z \in \mathbb{A}, z \in \mathbb{D}\}|}{|\{z: z \in \mathbb{D} \}|}
\end{equation*}
Since $a \times |\mathbb{A}|$ elements of set $\mathbb{D}$ came from set $\mathbb{A}$,
\begin{equation}\label{eq:3}
PC = \frac{(a \times |\mathbb{A}|)}{|\mathbb{D}|}  
\end{equation}

Now, using equation~\ref{eq:1},
\begin{eqnarray}
PC &= \frac{|\mathbb{D}| - b \times |\mathbb{B}|}{|\mathbb{D}|}\nonumber 
\\ &= 1 - b \times \frac{|\mathbb{B}|}{|\mathbb{D}|} \label{eq:4}
\end{eqnarray}
 
Now, the ratios $\frac{|\mathbb{B}|}{N}$ and $\frac{|\mathbb{D}|}{N}$ are essentially $P(Y = 1| X \leftarrow 0)$ and $P(Y = 1| X \leftarrow 1)$ (as per the notation used by Dawid), respectively. Therefore: 

\begin{equation}
\frac{|\mathbb{D}|/N}{|\mathbb{B}|/N} = \frac{P(Y = 1| X \leftarrow 1)}{P(Y = 1| X \leftarrow 0)} = RR \label{eq:5}
\end{equation}

where RR indicates the Risk Ratio. This means, using Equation~\ref{eq:4} the value of PC becomes:

\begin{equation}\label{eq:6}
 PC = 1 - b \times \frac{1}{RR} 
\end{equation}

\section*{Upper and Lower Limits of PC:}
Conceptually, PC will reach its highest value when all the elements
of set $\mathbb{D}$ come from set $\mathbb{A}$, and its lowest value
when none of the elements of set $\mathbb{D}$ come from set
$\mathbb{A}$. In general, the two conditions correspond to the value
of the fraction $b$ being $0$ and $1$, respectively. So, using
Equation \ref{eq:4}, we get the highest and lowest values of PC to
be $1$ and $(1 - \frac{1}{RR})$.  

Now, there are two special cases that limit PC to stricter bounds.
\begin{itemize}
\item \textbf{Case 1:} When $|\mathbb{A}| < |\mathbb{D}|$, even if all elements from set $\mathbb{A}$ go to set $\mathbb{D}$, some elements of set $\mathbb{D}$ still would have come from set $\mathbb{B}$. Which means under this condition, even with $a = 1$, $b$ can't really have a minimum value of 0, and the maximum value of PC therefore would be limited to $\frac{|\mathbb{A}|}{|\mathbb{D}|}$. So, under this condition, substituting $a=1$ in Equation \ref{eq:3}, we get:
$$
PC_{max} = \frac{|\mathbb{A}|/N}{|\mathbb{D}|/N} = \frac{P(Y = 0 | X \leftarrow 0)}{P(Y = 1 | X \leftarrow 1)}
$$

\item \textbf{Case 2:} Conversely, when $|\mathbb{B}| > |\mathbb{D}|$, even when $a = 0$, $b$ can't reach the maximum value of 1. 
So, substituting $a=0$ in Equation~\ref{eq:3}, we get $PC_{min} = 0$.
\end{itemize}

Therefore, we conclude:
$$
min\{1, \frac{P(Y = 0 | X \leftarrow 0)}{P(Y = 1 | X \leftarrow 1)}\} \geqslant PC \geqslant max\{0, (1 - \frac{1}{RR})\}
$$

\section*{Getting a distribution for PC:}
Using Equations~\ref{eq:3} and~\ref{eq:4}, we can say that if we can
get a distribution over the fractions $a$ and/or $b$, we can get a
distribution for PC as well. A few methods that can be used for
getting the distribution over $a$ and $b$ are listed below: 

\begin{itemize}
\item \textbf{Data from multiple experiments:} For a one-to-one mapping scenario, as described above, we can use data from multiple experiments, each of which should give a slightly different estimate for $a$ and $b$, which can, therefore be used to obtain a distribution over the fractions.
\item \textbf{Bootstrapping:} Another way of obtaining the distribution is using multiple bootstrapped sampled datasets to estimate $a$ and $b$.
\item \textbf{Different matching functions:} Choosing different matching functions would potentially alter the matching pairs, giving rise to a different mapping, therefore giving different estimates for $a$ and $b$, which can be used to obtain a distribution.
\item \textbf{Using a set of matching elements:} If set of matching
  elements is used instead of only one, it is possible to vary the number of elements considered and thereby obtain different estimates for $a$ and $b$, which can give us a distribution.
\item \textbf{Sampling:} For almost all datasets containing
  observational data, the dataset that is used for calculating PC is
  constructed by sampling from the dataset. By constructing multiple
  calculation datasets by sampling the original dataset, it is
  possible to get a distribution over PC. This is, perhaps, the most
  reliable way of obtaining the distribution, since the uncertainty
  is generated from the original dataset, and isn't an artifact of
  the method or function used. 
\end{itemize}

\section*{Corollary 1: Value of PC under Monotonicity Assumption}
Under the assumption of Monotonicity, as defined by
Pearl~\cite{tian2000probabilities}, we would have a condition that
no element from Set $\mathbb{B}$ would go to Set $\mathbb{C}$ (note
that this would be impossible if $|\mathbb{B}| > |\mathbb{D}|$);
which implies that all elements of $\mathbb{B}$ would go to Set
$\mathbb{D}$, which in turn would mean that the fraction $b$ would
be 1, making the value of PC a single value: $(1 - \frac{1}{RR})$. 

\section*{Corollary 2: Value of PC under Reverse-Monotonicity Assumption}
This is a case not discussed by Pearl or Dawid. Under this assumption, all elements of Set $\mathbb{B}$ would go to Set $\mathbb{C}$ (note that this would be impossible if $|\mathbb{B}| > |\mathbb{C}|$). Referring to the Aspirin trial example given by Dawid~\cite{dawid2016statistical}, this condition would mean if Ann does not have headache to begin with, then after taking aspirin, she will definitely get headache. Now this scenario may not be very common for the Aspirin trial case, because it essentially goes against the intended course of action of the medicine, but there could be other situations where it can be regarded as a possibility. This condition would force the value of the fraction $b$ to be 0, therefore PC will take the value of 1.

Note that, under this condition PC can not be less than 1, since all elements of Set $\mathbb{B}$ would go to Set $\mathbb{C}$, which would mean all elements of Set $\mathbb{D}$ would come from Set $\mathbb{A}$, making PC = 1.

\section*{Corollary 3: PC formula if a set of matching elements are selected}
If instead of selecting the one closest matching element we choose to select multiple matching elements (for each element in Set $\mathbb{D}$), as mentioned while describing the mechanism, the PC formula will be slightly altered. Let's assume for every element $z$ in $\mathbb{D}$, we choose $M$ closest matching elements, defined by a set $Z^*$ ($|Z^*| = M, Z^* \subseteq \{\mathbb{A},\mathbb{B}\}$). The formula for PC would then become:
\begin{equation}\label{eq:pcp}
PC = \frac{\sum_{z \in \mathbb{D}} ( |\{i : i \in Z^*, i \in \mathbb{A}\}|/M ) }{|\mathbb{D}|}
\end{equation}

The PC formula described in Equation \ref{eq:pc} is actually a special case of this  when $Z^*$ consists of only one element $z'$.

\section*{Comparing the result with the formulas given by Dawid and Pearl}
The bounds we obtained for the value of PC is exactly the same as shown by Dawid~\cite{dawid2016statistical,dawid2016bounding}. Pearl uses a different notation, but the values are exactly the same under specific conditions, as was mentioned in \cite{tian2000probabilities}.

\section*{Addressing the limitations of the approaches by Pearl and Dawid}
Our definition of probability is also not directly based on
likelihood of an event. Instead, our approach focuses on finding the
closest matching pairs of elements, therefore, the ``event" in this
context would indicate the matching of elements. However, this
definition, while not be exactly similar to the original definition
of probability, still has a clearer mechanism as compared to the
counterfactual definition, and is able to address the limitations of
the approaches taken by Pearl and Dawid, as listed below: 
\begin{itemize}
\item \textbf{Addressing the problems with counterfactual
    philosophy:} As mentioned above, our definition of PC does not
  try to estimate the value of PC by assuming an alternate reality
  for an observed element, instead focusing on the behavior of the
  closest matching elements under the different realities. 
\item \textbf{Obtaining a distribution for PC:} The way our method
  can be used to obtain a distribution for PC is described above in
  detail. 
\item \textbf{Introducing Uncertainty:} The uncertainty is
  introduced in our formula primarily by using different matching
  functions to choose different matching elements for a given
  element. If a set of elements is used instead, further uncertainty
  is introduced by choosing how many matching elements to use. This
  uncertainty is translated to the estimates of $a$ and $b$, and
  thereby in the final estimate of PC. 
\item \textbf{The G2i problem:} The way our method handles the G2i
  problem, with or without the exchangeability assumption, is
  mentioned above in detail. 
\end{itemize}


\section{Robustness Issues in calculating PC using the Existing Methods}\label{s:method}

Another problem we faced while applying the PC formula to practical situations was its sensitivity to data errors. From the definition of PC, if $P(Y = 1| X \leftarrow 1)$, or $P(X =1, Y=1)$ in general, is very small, the value of PC would be more sensitive to small changes in data. This problem is amplified a lot if Pearl's PN formula is used to calculate the value of PC from two datasets, one experimental and one observational, as illustrated in \cite{tian2000probabilities}. Small measurement errors are very common in almost all real life applications, so, if the value of PC changes a lot for small changes in data, it can not be used reliably. 

\subsection*{Robustness Issue: Example}

\begin{table}[h]
\caption{Example dataset from Tian \& Pearl~\cite{tian2000probabilities}, showing deaths and survivals among users (x) of a certain hypothetical drug  and drug non-users (x$'$)}
\label{t:pearl1}
\begin{tabular}{l|l l|l l|}
\cline{2-5}
                                     & \multicolumn{2}{l|}{Experimental} & \multicolumn{2}{l|}{Non-experimental} \\ \cline{2-5} 
                                     & x               & x$'$            & x                 & x$'$              \\ \hline
\multicolumn{1}{|l|}{Deaths (y)}     & 16              & 14              & 2                 & 28                \\ \hline
\multicolumn{1}{|l|}{Survivals (y')} & 984             & 986             & 998               & 972               \\ \hline
\end{tabular}
\end{table}

Let us take the example given in \cite{tian2000probabilities} (shown
in Table \ref{t:pearl1}) for illustrating the robustness issues. If
we use Pearl's PN formula on this dataset, we get the value of PN
(which is same as PC) to be 1. Now death of a person is a stable
situation, which very little room for error, unlike, for example,
blood glucose level or body temperature for a patient, which are
more susceptible to measurement errors. But even for this example,
assume a hypothetical situation where a second group of scientists
used the same dataset, but instead of checking if the patients died
within, for example, 7 days after getting the drug under
experimental condition as was examined by the first group, they
recorded if the patients had died within 10 days. Let's assume they
found one more death among the drug non-users, making the number 15,
instead of 14. This is a change of 1 in 1000, so one would assume it
should not change the result too much. But if we apply Pearl's PN
formula, we find that the value of PN has now become 0. So, a change
of 1 in 1000 in only one of the 4 conditions gives us a completely
opposite picture as to whether or not the drug was the cause of
death. Such a drastic change is not acceptable while applying the
formula in any real life situation.

\begin{figure}
\includegraphics[width = \linewidth]{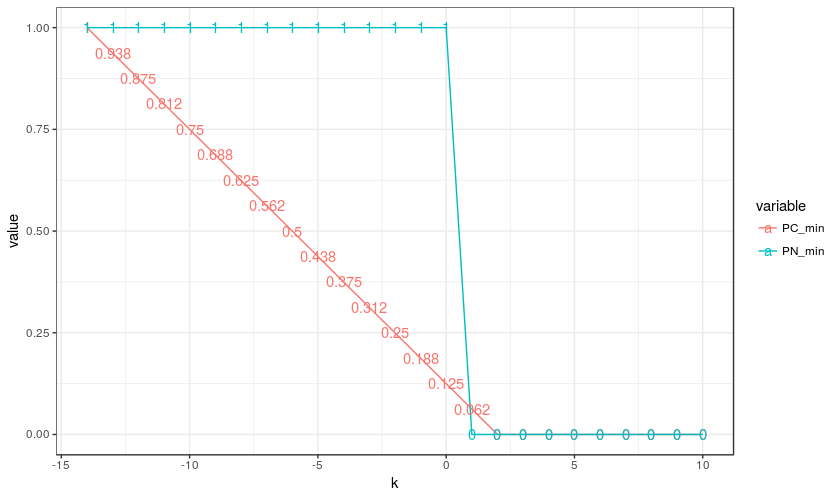}
\caption{Variation of PC and PN values with changes to data - case 1}
\label{fig:pc_pn1}
\end{figure}
If the value of PC is calculated only from the observational data, however, the rate of change is much slower, due in part to the fact that P(x,y) is much smaller for the Non-experimental data. If we change the value under experimental (x$'$,y) cell from 14 to (14+k), Figure \ref{fig:pc_pn1} shows how the value of Probability of causation (the minimum limit) changes with k for PN, calculated from both tables with Pearl's formula, and PC, calculated only from the experimental table. 

\begin{table}[h]
\caption{Example dataset from Dawid~\cite{dawid2016bounding}, showing deaths and survivals among users (x) of a certain hypothetical drug  and drug non-users (x$'$)}
\label{t:dawid1}
\begin{tabular}{l|l l|l l|}
\cline{2-5}
                                     & \multicolumn{2}{l|}{Experimental} & \multicolumn{2}{l|}{Non-experimental} \\ \cline{2-5} 
                                     & x               & x$'$            & x                 & x$'$              \\ \hline
\multicolumn{1}{|l|}{Deaths (y)}     & 30              & 12              & 18                & 24                \\ \hline
\multicolumn{1}{|l|}{Survivals (y')} & 70              & 88              & 82                & 76               \\ \hline
\end{tabular}
\end{table}

\begin{figure}
\includegraphics[width = \linewidth]{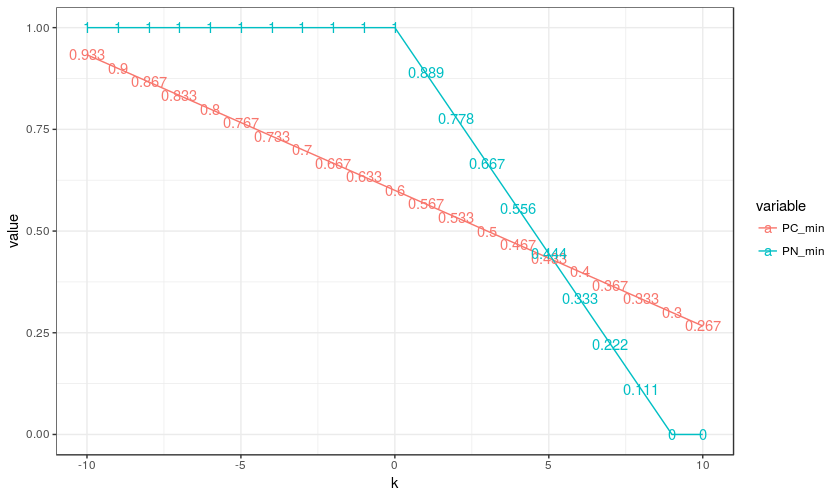}
\caption{Variation of PC and PN values with changes to data - case 2}
\label{fig:pc_pn2}
\end{figure}

As a second example, we take the dataset given by Dawid in \cite{dawid2016bounding}, shown in Table \ref{t:dawid1}. In this case we do not as drastic a change as in the last case, but still PN changes from 1 to 0 in 9 observations, which is still less than 10\% of the total observations. Direct calculation of PC is again found to to be more stable. The result of changing the value of experimental (x$'$,y) cell from 12 to (12+k) is shown in Figure \ref{fig:pc_pn2}.

\subsection*{Proposed Improvement using our method}
By using our proposed method, it is possible to get a distribution over the value of PC. Moreover, since our method is more akin to the PC formula used by Dawid, it would be possible to avoid such drastic change in the value of PC. We could not use the same examples for illustration, because as mentioned before, we need additional data for matching different observations. However, we give an illustrative example showing how our method works in the following section.

\section{Examples}\label{s:example}

In this section we provide two illustrative examples of application of our method of calculating PC. The first example is based on a very simple custom situation, which is also used to illustrate the effects of sample size, rare events, and matching 1, 3, or 5 closest matching samples. The second example is using the LUCAS dataset\footnote{http://www.causality.inf.ethz.ch/data/LUCAS.html}, examining the probability that smoking was the cause of lung cancer for the patients who smoked and developed lung cancer. 

\subsection{Example 1: Custom generated data}
In this example we have a hypothetical scenario where the cause event is application of a medicine, and the effect event is death resulting from its application. We use only one variable called ``Id" for matching, and we use a identity function as the matching function, which returns 1 if the ``Id'' matches (for one-to-one matching, otherwise it returns 1 if the difference of ``Id" is less than a certain threshold). Besides illustrating our method, we use this scenario to examine the stability of the distribution of PC for different sample sizes, for rare events (X=1, Y=1 is a rare event), and for matching 1, 3, or 5 closest matching elements. For this scenario, we intentionally keep the ratio of cardinalities of sets $\mathbb{A}$ and $\mathbb{B}$ as 8:2, so the value of PC should be $0.8$, since we generated the ``Id" values from a uniform distribution. By definition, the range of values for PC should lie between 0.5 and 1 in this situation.

For both the cases, ``P1" indicates the situation where the mapping was one-to-one, and ``P3" and ``P5" indicate the situation where, respectively,  3 and 5 closest  elements were matched with every element of Set $\mathbb{D}$.

\subsubsection*{Effect of sample size:}
First, we decided to see how the resultant distribution for PC varies with sample size (N). 
To examine that, we ran 1000 iterations by randomly generating the ``Id'' variable in each iteration, for the following values of N: 5, 10, 50, 100, 500, 1000, 5000. The ratio of cardinalities of sets $\mathbb{C}$ and $\mathbb{D}$ were kept 6:4 in these situations. 

\begin{figure}
\includegraphics[width = \linewidth]{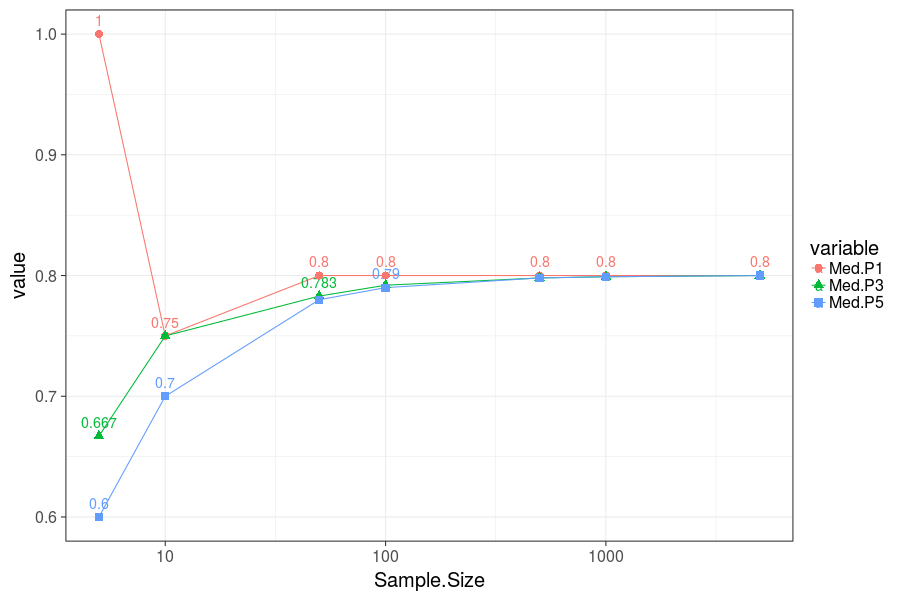}
\caption{Median value of PC for different Sample Sizes}
\label{fig:z_med}
\end{figure}

\begin{figure}
\includegraphics[width = \linewidth]{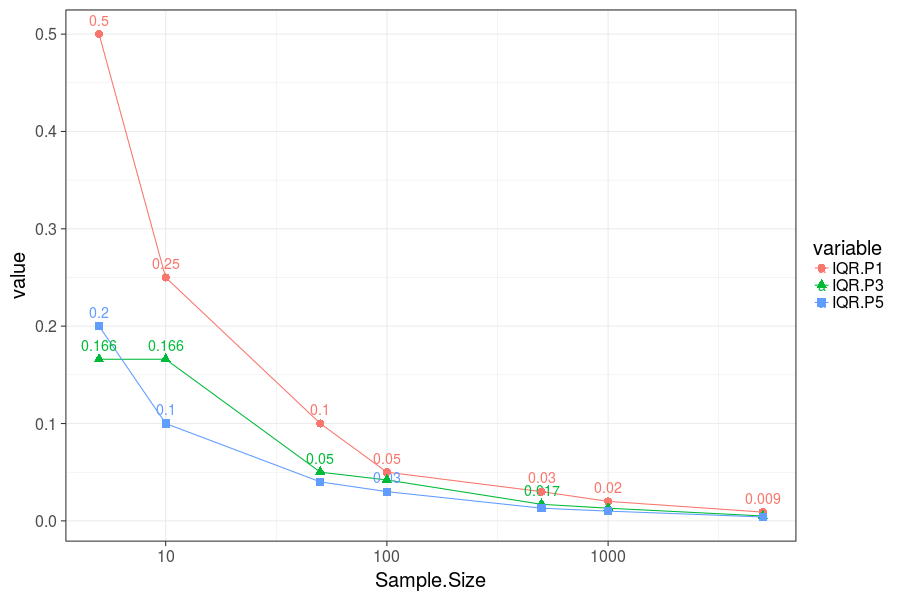}
\caption{Interquartile Range of PC for different Sample Sizes}
\label{fig:z_iqr}
\end{figure}

\begin{figure}
\includegraphics[width = \linewidth]{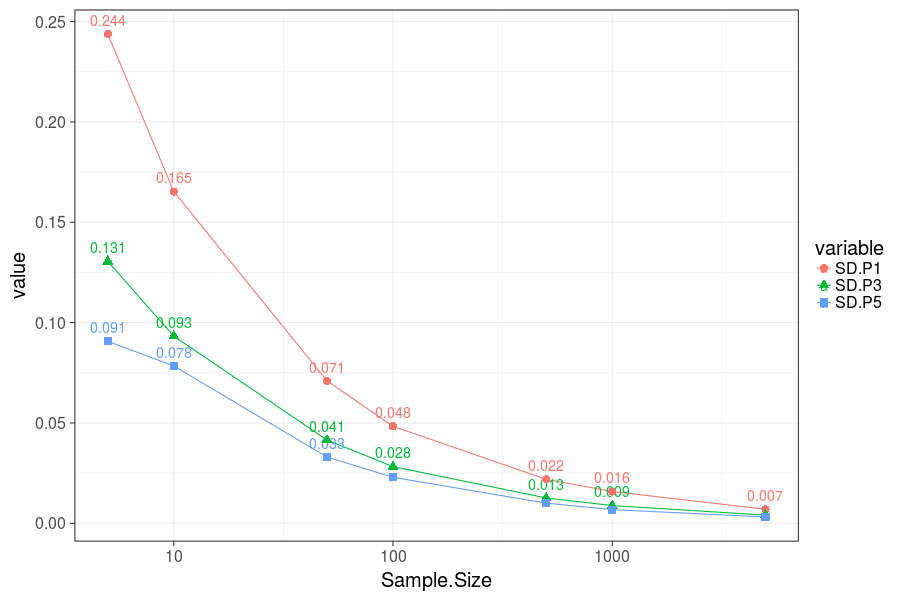}
\caption{Standard Deviation of PC for different Sample Sizes}
\label{fig:z_sd}
\end{figure}

It can be seen from Figure~\ref{fig:z_med} that if the sample size is very low, then the PC value we get tends to be inaccurate. When the sample size goes above 100, we become very close to the correct value, and for more than 500, we always have the correct value. Matching more elements seems to require a larger sample size to converge to the correct value. However, we find from Figures~\ref{fig:z_iqr} and~\ref{fig:z_sd} that matching more elements consistently result in lower values of interquartile range and standard deviation, resulting in tighter distributions.

\subsubsection*{Effect of Rare Event:}
By event, we mean the observed event, which is the situation X=1, Y=1 in this context. In such situations, getting a believable distribution is difficult, since there are very few elements to match. This also highlights how our method works for situations like the ones described in Section~\ref{s:method}. As mentioned before, if both observational and experimental data are available, we suggest matching the cases in the observational setting for X=1, Y=1 with the cases in experimental setting for X=1, Y=1, with sampling if applicable, and use the modified dataset for further calculation. A more concrete example with sampling is presented in the next example. 

In this situation, we use a fixed sample size of 1000, and vary the ratio of cardinalities of Sets $\mathbb{C}$ and $\mathbb{D}$ as $(1-r) : r$. The values of $r$ that we used are: 0.002, 0.005, 0.01, 0.015, 0.02, 0.03, 0.04, 0.05, 0.075, 0.1, 0.2, and 0.3, which correspond to Set $\mathbb{D}$ having 2, 5, 10, 15, 20, 30, 40, 50, 75, 100, 200, and 300 elements. This exercise was also repeated 1000 times.

\begin{figure}
\includegraphics[width = \linewidth]{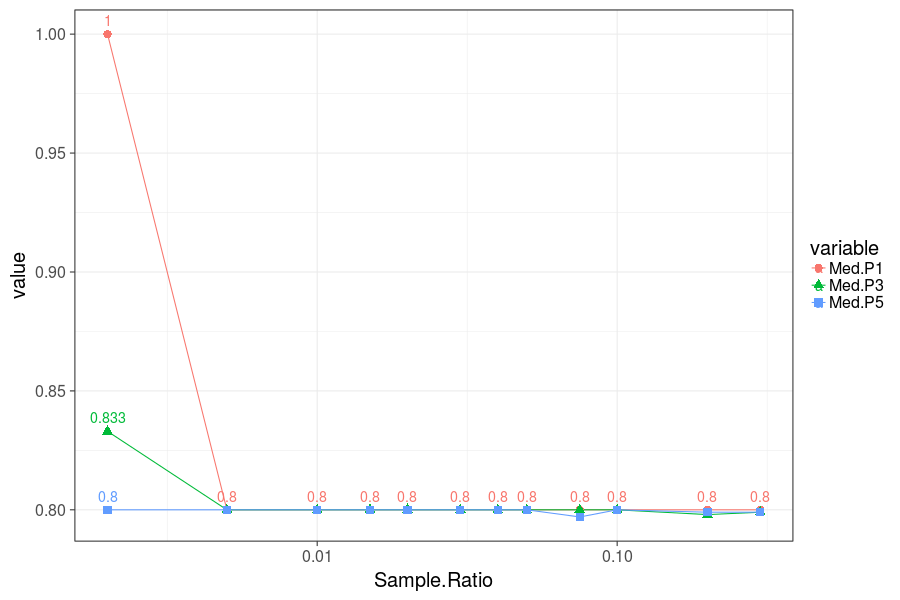}
\caption{Median value of PC for different Sample Ratio values}
\label{fig:s_med}
\end{figure}
\begin{figure}
\includegraphics[width = \linewidth]{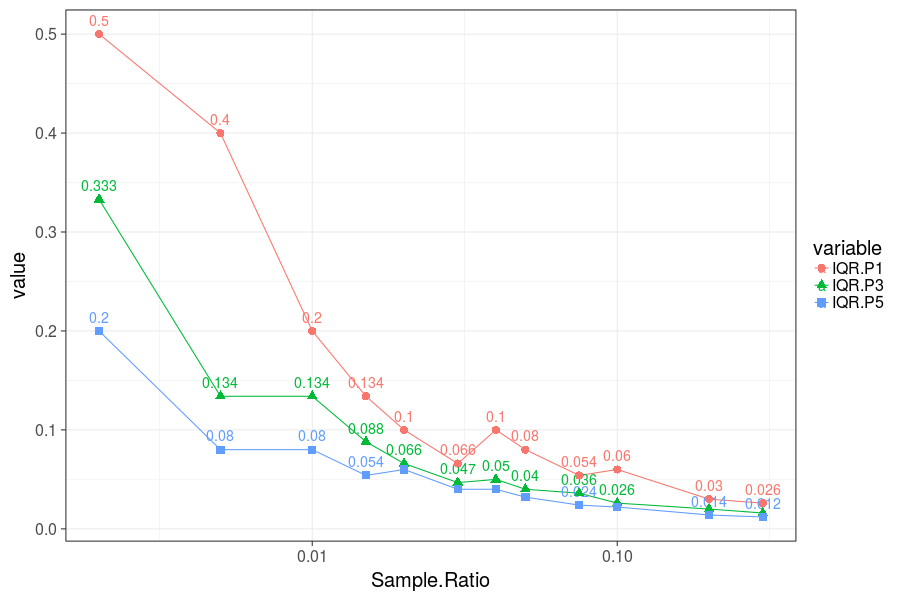}
\caption{Interquartile Range of PC for different Sample Ratio values}
\label{fig:s_iqr}
\end{figure}
\begin{figure}
\includegraphics[width = \linewidth]{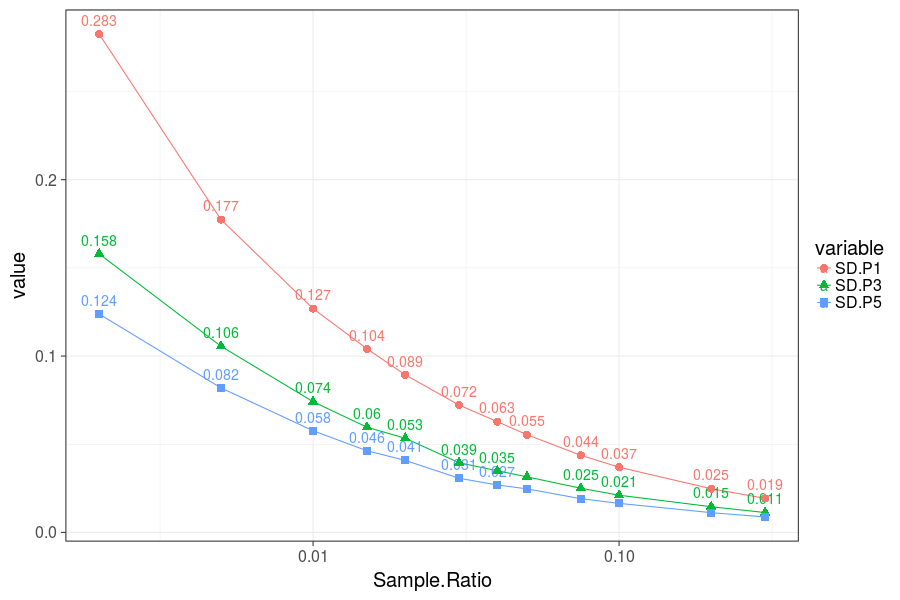}
\caption{Standard Deviation of PC for different Sample Ratio values}
\label{fig:s_sd}
\end{figure}

As can be seen from Figure~\ref{fig:s_med}, when 5 closest elements are matched, even with only 2 observations in set $\mathbb{D}$, we obtain the correct value of PC, and for other cases, it settles to the correct value for only 5 elements in Set $\mathbb{D}$. The little bit of wiggle for larger values of sample ratio are likely random in nature. 

We also notice that, as expected, the interquartile range and standard deviation of PC are much larger for rare events, as can be seen from Figures~\ref{fig:s_iqr} and~\ref{fig:s_sd}. However, having more matching elements is shown be constantly giving a more stable result in all cases.

\subsubsection*{Conclusion from the Results}
In this example we illustrate how our method can be used for calculating the value as well as the distribution for PC. The results indicate that a sample size of more than 100 (around 500 or more for better accuracy) is needed to get an accurate measurement, and matching each element in Set $\mathbb{D}$ with multiple elements in Sets $\mathbb{A}$ and $\mathbb{B}$ results in a tighter distribution. When the observed event is rare, our method can still get an accurate estimate of PC for as few as 5 observations, and in this case also, matching multiple elements produce better results. 

\subsection{Example 2: Using LUCAS dataset}

\begin{figure}
\includegraphics[width = \linewidth]{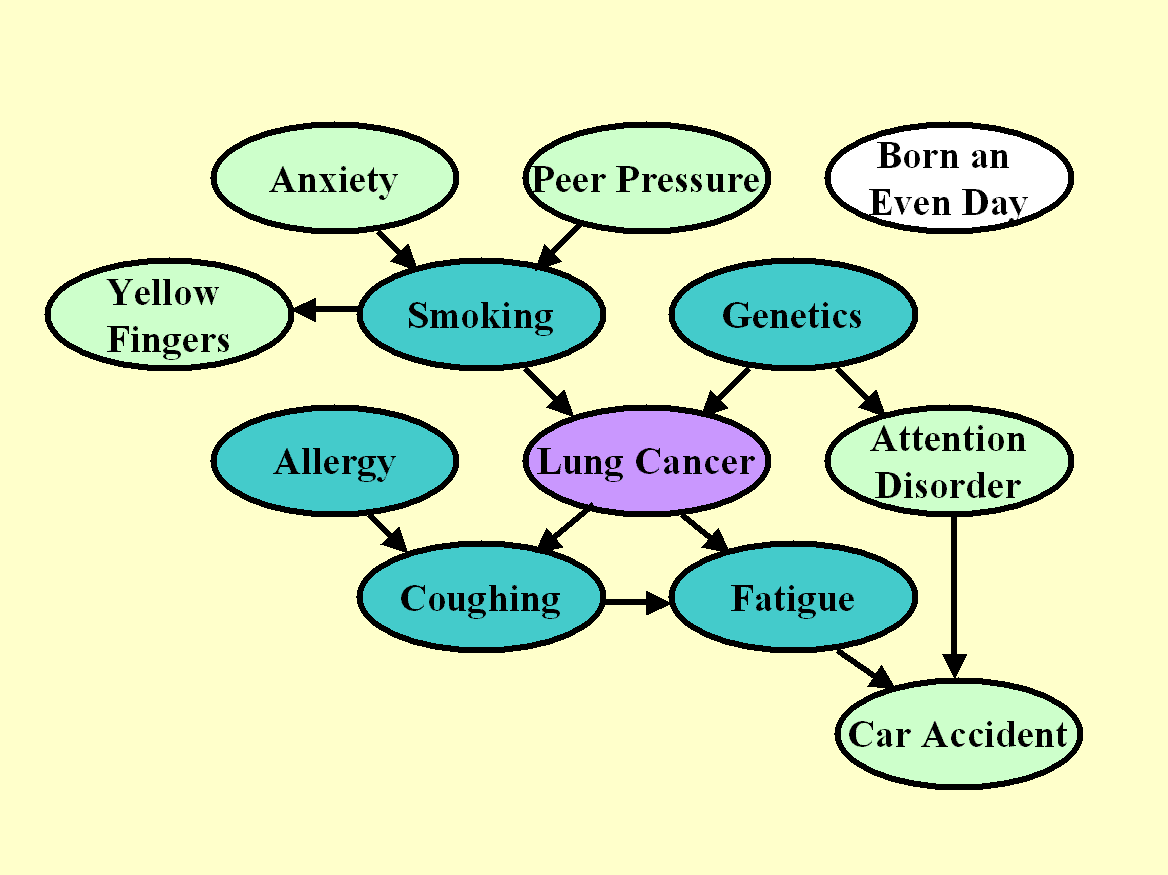}
\caption{Bayesian Network for LUCAS dataset}
\label{fig:lucas_bn}
\end{figure}

Our second example is based on the LUCAS (LUng CAncer Simple set) dataset\footnote{http://www.causality.inf.ethz.ch/data/LUCAS.html}, which contains data generated artificially by causal Bayesian networks with binary variables. The Bayesian network for this dataset is shown in Figure~\ref{fig:lucas_bn}.

In this example we examine the link between ``Smoking'' and ``Lung Cancer''. Specifically, the question we are interested in answering is that for the individuals who were smokers and had lung cancer, what is the probability that smoking was the cause of cancer. From the causal diagram we can see that ``Genetics'' is another factor that has in influence over ``Lung Cancer'', so the question of probability of causation is a valid question.

Since this is not an experimental dataset, the number of observations for our two conditions: smoking and non-smoking are not same. Out of the 2000 observations in the dataset, 1505 are for smokers and 495 are for non-smokers. Note that, for our formula to work, we need equal number of observations for the two conditions. So, we decided to use sampling to create the dataset we used for further analysis, which we call the analysis dataset. Based on the result of the earlier example, and the number of observations for the two conditions, we decided to create an analysis dataset with 400 observations for smoking and non-smoking conditions. 

The default way of sampling that we use takes 400 random samples from the set of smokers, and 400 from the set of non-smokers, and calculate PC from that dataset. However, we suggest that if the demographic data about what percent of smokers get cancer and what percent of non-smokers get cancer is available, we suggest that information should be used for a more accurate estimate. For example, if we know from other sources that 60\% of smokers get cancer and 25\% of non-smokers get cancer, then we suggest that 240 (60\% of 400) samples should have been taken from the set of smokers who got cancer, 160 (40\% of 400 ) from the set of smokers who didn't get cancer, 100 (25\% of 400) from the set of non-smokers who got cancer, 300 (75\% of 400) from the set of non-smokers who didn't get cancer. Adding this additional information should result in a tighter and more accurate distribution of PC. Otherwise, if there is reason to believe that the observational data is a faithful representation of the distribution of smoker cancer patients and non-smoker cancer patients, the ratio from the observed data can also be used. 

In our example, we looked at both the options: the default option of sampling without caring about the ratio, and the other option where we assume the dataset is a faithful representation of the distribution of smoker cancer patients and non-smoker cancer patients, and use the ratios observed from the dataset. We ran this exercise 1000 times for both the options, creating different analysis datasets in each step, and used the same matching function ( from the MatchIt~\cite{rmi} library in R ). So, the distribution we obtain for PC is an artifact of the dataset and not an artificial artifact generated by the matching function.

\subsection*{Option 1: Direct sampling from the dataset}
When we used a direct sampling approach from the Lucas dataset, we found that the value of PC ranges from 0.6100 to 0.7118, with a median of 0.6569 and standard deviation of 0.0143. Using the upper and lower bound formulas, we found that for the 1000 iterations, the value of PC would range between 0.5952 and 0.7857. 

\subsection*{Option 2: Sampling by conserving the ratio from the dataset}
For this option we assume that the ratio of smoker cancer patients and non-smoker cancer patients observed in the dataset is representative of the ratios in the demographic population under consideration. 
In the original dataset, we have 328 non-smokers who didn't have cancer, 167 non-smokers who got cancer, 229 smokers who didn't have cancer, and 1276 smokers who got cancer. So, for non-smokers, we take the ratio of cancer patients and individuals without cancer to be 1:2, and for smokers the ratio is taken to be 11:2. 

With this constraint in place, we found that the value of PC ranges from 0.6331 to 0.6982, with a median of 0.6627 and standard deviation of 0.0101. Using the upper and lower bound formulas, we found that for the 1000 iterations, the value of PC would range between 0.61 and 0.79. So, adding this constraint results in a tighter bound for PC.

\subsubsection*{Conclusion from the Results}
We can see that our method produces a distribution for PC which is a more stable estimate than just the theoretical minimum and maximum values. In this example we show how our method can be used estimate PC from purely observational data. We also show how demographic information from other sources can be used to get a tighter estimate of PC using our method.

If the goal is to estimate the value of PC for a single individual (presumably a smoker cancer patient in this case), we suggest using a matching algorithm and from the original dataset select only the observations for which the value of the matching function is above a certain threshold. We suggest the threshold should be chosen subjectively depending on the particular scenario. Then our method can be used to estimate PC from this reduced set. This reduction is used to eliminate the cases which are very dissimilar to the case under consideration, so we suggest using a relatively low value for the threshold so that there are enough observations to get a stable distribution for PC. The theory for this is mentioned in the remarks following Equation~\ref{eq:pc}.

\section{Limitations}\label{s:lim}

Although we believe our approach offers simpler and richer way of estimating the value of PC, there are a few limitations to our approaches, as listed below:
\begin{itemize}
\item \textbf{Need of a richer dataset:} Our approach can give the bounds of the value of PC from an aggregate dataset, same as the methods used by Pearl and Dawid, but for obtaining the extra informations, a richer dataset needed. Basically, for each individual element in question, we need some extra information on them that can be used for finding matches. However, we believe that obtaining such a dataset is not an issue for most of the present day situations.
\item \textbf{Finding the right matching function:} Finding the appropriate matching function is another requirement, which is a non-trivial question. Therefore, we recommend trying a set of possible matching functions for obtaining a more reliable result.
\item \textbf{The effect of hidden bias on matching:} Hidden bias may actually increase because matching on observed variables may unleash bias due to dormant unobserved confounders. 
Similarly, Pearl~\cite{pearl2009causality} has argued that bias reduction can only be assured (asymptotically) by modeling the qualitative causal relationships between treatment, outcome, observed and unobserved covariates.

\end{itemize}


\section{Conclusion}\label{s:conclusion}

In this paper we have presented a methodology for estimating the distribution for Probability of Causation using matching. Our method takes a set theory based approach to identify the different scenarios and matching for getting a distribution for PC. 

The limiting values of PC from our method are same as what was proposed by Pearl and Dawid, but where the previous works could only offer the limiting values, our method can get a distribution for PC, and therefore should be more useful in practice. Moreover, our method can be used on purely observational dataset, which should be immensely useful in practice since performing an unbiased experiment is difficult, if not impossible in many practical scenarios. 





\bibliographystyle{model1-num-names}
\bibliography{sample.bib}







\end{document}